\documentclass[prl,twocolumn,aps,amsmath,superscriptaddress,showpacs]{revtex4}

\usepackage{graphicx}
\usepackage{amssymb}

\newcommand\Name[1]{{#1},}
\newcommand\REVIEW[4]{{\em{#1}} {\bf{#2}}, #4 (#3)}

\usepackage[usenames,dvipsnames]{xcolor}

\begin{document}

\title{Heating without heat: thermodynamics of passive energy filters between finite systems}

\author{R.~Mu\~noz-Tapia}
\affiliation{F\'{\i}sica Te\`orica: Informaci\'o i Fen\'omens Qu\`antics, Departament de F\'{\i}sica,
Universitat Aut\'onoma de Barcelona, 08193 Bellaterra (Barcelona), Spain}

\author{R.~Brito}
\affiliation{ Departamento de F{\'i}sica Aplicada I and {\em GISC},
Universidad Complutense de Madrid, 28040-Madrid, Spain}

\author{J.M.R.~Parrondo}
\affiliation{Departamento de F{\'i}sica At{\'o}mica, Molecular y
Nuclear and {\em GISC}, Universidad Complutense de Madrid,
28040-Madrid, Spain}

       %% Nonequilibrium and irreversible thermodynamics.

\begin{abstract}
Passive filters allowing the exchange of particles in a narrow band of
energy are currently used in micro-refrigerators and energy
transducers. In this letter, we analyze their thermal properties
using linear irreversible thermodynamics and kinetic theory, and discuss a striking
phenomenon: the possibility of increasing or decreasing simultaneously the
temperatures of two systems  without any supply of energy. This occurs when the filter induces a flow of particles whose energy is between the average energies of the two systems. 
Here we show that this selective transfer of particles does not need the action of any sort of Maxwell demon and can be carried out by passive filters without compromising the second law of thermodynamics. The phenomenon allows us to design cycles between two reservoirs at temperatures $T_1<T_2$  that are able to reach temperatures below $T_1$ or above $T_2$.

\end{abstract}

\maketitle

Our knowledge on the transfer of energy between physical systems has experienced a considerable growth in the last years. 
New fields like stochastic \cite{stochtherm}Â  and quantum 
thermodynamics \cite{qtherm} Â extend the concepts of heat and work to fluctuating microscopic systems both in the classical and the quantum regime.
These recent developments are partly driven by the possibility to manipulate microscopic systems, such as colloidal particles in optical traps or  single electron boxes. A technical achievement with profound consequences on thermodynamics is the construction of passive energy filters that allow the selective transfer of particles with a specific energy.

Bandpass energy filters for ballistic electrons based on resonant tunneling were introduced in
the 90's and can be implemented in semiconductor super-lattices
and  quantum dots \cite{capasso}, nanowires \cite{nanowires,boukai}, etc. Some thermodynamic
effects of such filters have been already studied: Pekola  used filters between semiconductors and/or superconductors to build up
novel cooling mechanisms \cite{pekola}. Humphrey and
Linke  designed { energy selective electron heat
engines} based on Brownian ratchets \cite{linke}. Filtering resonant nano-structures have been proposed to enhance the efficiency of thermoelectric devices \cite{linketherm,chris,linketherm2,boukai}. Other combinations of filters and non-equilibrium sources, like hot photonic reservoirs  \cite{chris1} or ac voltages \cite{sols}, have  been also explored to find novel and interesting thermal phenomena with potential applications. From the point of view of linear irreversible thermodynamics, filters induce tight coupled fluxes that reach high efficiencies in generic energy transduction setups \cite{chris05,casati08,izumida14}.

Nevertheless, none of these previous works explored the
thermodynamic consequences of selective particle exchange between
{\em finite} systems, i.e., systems which are not reservoirs and whose temperature and density is affected by the flow of particles \cite{finiteness}. It is in this context that one can find 
effects which, at first sight, seem to defeat fundamental thermodynamic limitations.

Here we show 
one of those effects, the simultaneous increase of temperature of two 
systems without any external energy supply.
In other words, the possibility of ``heating without heat''. The opposite, i.e., 
spontaneous simultaneous cooling, can also occur. The reason behind these striking behaviors is that temperature is related to the average energy, that depends both on the total energy and the total number of particles. As in evaporative cooling, temperature can decrease (increase) if particles with high (low) enough energy leave the system \cite{evaporative}. 
The combination of this effect  and the selective exchange of particles induces a rich and unexpected phenomenology. 

The aim of this Letter is first to formulate a complete and consistent theoretical framework for the thermodynamics of finite width energy filters connecting arbitrary reservoirs and, secondly, to explore the aforementioned phenomena.

We start by using standard thermodynamics to characterize the equilibrium
state of two systems, 1 and 2, separated by an adiabatic wall with an ideal filter 
that allows the exchange of particles with energy $E_{\rm f}$. Since the only transfer of energy
between the two systems is due to the exchange of particles, the
following constraint holds:
\begin{equation}
dE_i=E_{\rm f}\,dN_i\qquad i=1,2
\label{constraint}
\end{equation}
where $E_i$ is the internal energy and $N_i$ the number of particles
in each system. The global system is described by four variables,
$N_1,N_2,E_1,E_2$, but, due to conservation of total energy
($dE_1=-dE_2$) and particles ($dN_1=-dN_2$), together
with (\ref{constraint}), only one of those is independent. Choosing $N_1$ as
independent variable and using the three constraints, the
entropy differential reads \cite{linke}:
\begin{eqnarray} dS 
&=& \left[ \frac{E_{\rm f}-\mu_1}{T_1} -
\frac{E_{\rm f}-\mu_2}{T_2}\right] d N_1=\Delta\alpha\, d N_1
\label{entropy}
\end{eqnarray}
where $T_i$ are the respective temperatures of the two systems,
$\mu_i$ the chemical potentials and $\Delta \alpha =\alpha_1-\alpha_2$, with $\alpha_i=(E_{\rm f}-\mu_i)/T_i$,
is the thermodynamic force conjugate to the flow of particles.
The resulting equilibrium condition is $\Delta\alpha=0$, which
does not necessarily imply that temperatures and chemical potentials
are equal in both subsystems \cite{linke,linketherm,chris}. The corresponding phenomenological
equation for the flow of particles reads: 
\begin{equation} \dot N_1
=\kappa\Delta \alpha \label{pheno}
\end{equation}
where $\kappa$ is a transport coefficient which is positive to
ensure a positive entropy production ($\dot S=\kappa\Delta
\alpha^2\geq 0$).

This simple analysis predicts a non trivial equilibrium state, given by $\Delta \alpha=0$, 
and provides a dynamical equation for the relaxation towards that equilibrium state. However, the whole argument relies on meeting the constraint \eqref{constraint}, only valid for an ideal filter. Moreover, for such ideal filter one can suspect that the exchange of particles  is negligible, since the fraction of particles with an energy exactly equal to $E_{\rm f}$ is infinitesimally small. To clarify this issue we need to consider filters with a finite width $\Delta E$ 
 and perform a kinetic analysis of the exchange of particles.

Let us assume that the two systems are 
always in equilibrium with a well defined temperature $T_i$ and chemical potential $\mu_i$.
Then the number of particles $\phi_{i\to j}$ with energy in the interval $[E,E+dE]$ going from system $i$ to system $j$ per unit of
time only depends on the energy $E$ and the
thermodynamic state of system $i$:
$\phi_{i\to j}=\phi(E;T_i,\mu_i)dE$. 
Under this assumption, for a filter centered at $E_{\rm f}$ with width $\Delta E\to 0$, we have:
\begin{equation} \dot N_1=\left[\phi(E_{\rm f};T_2,\mu_2)-\phi(E_{\rm f};T_1,\mu_1)\right] \Delta E. \label{flows}
\end{equation}
Since $\dot N_1=0$ when $\Delta\alpha=0$ 
for any
value of $E_{\rm f}$, the flow $\phi(E_{\rm f},T,\mu)$ must be a function of $E_{\rm f}$ 
and $\alpha\equiv (E_{\rm f}-\mu)/T$, hence
$\phi(E;T,\mu)=f\left( E,(E-\mu)/T\right)$.

Consider now a narrow filter of finite width $\Delta E$, allowing the transfer of particles with energy between $E_{\rm f}-\Delta E/2$ and $E_{\rm f}+\Delta E/2$. 
The net flow
of particles and energy from system 2 to system 1 obeys the evolution equation:
\begin{eqnarray} \dot N_1 &=& \int_{E_{\rm f}-\Delta E/2}^{E_{\rm f}+\Delta E/2} dE\left[ f_2(E)-f_1(E)\right]
\label{eq1}\\
\dot E_1 &=& \int_{E_{\rm f}-\Delta E/2}^{E_{\rm f}+\Delta E/2} dE\, E\,\left[ f_2(E)-f_1(E)\right]\label{eq2}
\end{eqnarray}
with $f_i(E)=f(E,(E- \mu_i)/T_i)$.

For narrow filters, we can expand the functions $f_i(E)$ up to linear terms in $(E-E_{\rm f})$.
Inserting the expansion in Eqs.~(\ref{eq1},\ref{eq2}) we obtain, up  to  third-order terms in $\Delta E$,
\begin{eqnarray} \dot
N_1 &=& \left[ f_2(E_{\rm f})-f_1(E_{\rm f})\right]\Delta E 
\label{linearfinaln}
\\ &+& \frac{ 
f_2''(E_{\rm f})-f_1''(E_{\rm f})
}{24}\Delta E^3+\dots
\nonumber
\\
\dot E_1 &=&  E_{\rm f}\,\left[ f_2(E_{\rm f})-f_1(E_{\rm f})\right]\Delta E  
\label{linearfinale}\\ 
&+&
\frac{1}{24}\,\left[ 
2f'_2(E_{\rm f})+E_{\rm f}f_2''(E_{\rm f})-(2\leftrightarrow 1)
\right]\Delta E^3+\dots \nonumber
\end{eqnarray}
The first-order terms in Eqs.~\eqref{linearfinaln} and \eqref{linearfinale}  describe a relaxation towards $f_1(E_{\rm f})=f_2(E_{\rm f})$, i.e., towards $\Delta\alpha=0$, which reproduces, for $\Delta\alpha$ small, the phenomenological equation \eqref{pheno} with $\kappa=-\left.\partial_\alpha\right|_{\alpha =\alpha_1}  f(E_{\rm f},\alpha) \Delta E$. 
On the other hand, the third-order terms in (\ref{linearfinaln}-\ref{linearfinale}) correspond to a much slower relaxation.

The evolution behaviour can be summarized as follows: for $\Delta\alpha$ large, there is a comparatively fast relaxation towards $\Delta\alpha=0$ described by Eq.~\eqref{pheno}, with a decay time of order $1/\Delta E$. When $\Delta\alpha \lesssim \Delta E^2$, 
terms of order $\Delta E^3$  dominate, inducing a slower relaxation along the line $\Delta\alpha\simeq 0$ towards $f_1(E)=f_2(E)\,\,\forall E$, i.e., to full equilibrium, $T_1=T_2$ and $\mu_1=\mu_2$, with a decay time of order $1/\Delta E^3$.

 Equations~(\ref{eq1}-\ref{linearfinale}) and the discussion above are
completely general: the specific nature of the
subsystems and the filter enters into the equation through the function $f(E, (E-\mu)/T)$. Actually, to apply  the equilibrium condition $\Delta\alpha=0$ we only need  
the equation of state of
the systems, that is, $\mu$ and $T$ as functions of $N$ and $E$. For instance, for two dimensional ideal classical gases in a volume $V$, 
$kT=E/N$ and $\mu=kT\ln(N\Lambda^2/V)$, where $k$
is Boltzmann constant and $\Lambda=h/\sqrt{2\pi mkT}$ is the thermal wavelength. The equilibrium condition, $\Delta \alpha=0$,
in terms of the temperatures and the numbers of particles, can be written as:
\begin{equation}\label{eqgi} \frac{n_1}{kT_1}e^{-\frac{E_{\rm f}}{kT_1}}= \frac{n_2}{kT_2}e^{-\frac{E_{\rm f}}{kT_2}}
\end{equation}
where $n_{i}=N_i/V$ is the particle density in system $i$.  

The precise form of the particle flow $f(E, (E-\mu)/T)$ depends on the filter.
A  reasonable assumption is that the fraction $\gamma(E)$ of particles in the unit volume 
that traverse  the filter per unit time only depends  on $E$. 
Then the flow of particles leaving the system reads
\begin{equation}
f\left(E,\frac{E-\mu}{T}\right)=\gamma(E)\frac{n}{kT}e^{-\frac{E}{kT}}\label{f2d}.
\end{equation}
This is the case, e.g.
for an effusion filter, where  the flow of particles of mass $m$ through a hole of length $L$  is given by the effusion rate \cite{effusion} $\gamma_{\rm eff}(E)=\sqrt{{2EL^2}/{m}}$.

\begin{figure}
\begin{center}
\includegraphics[height=7cm]{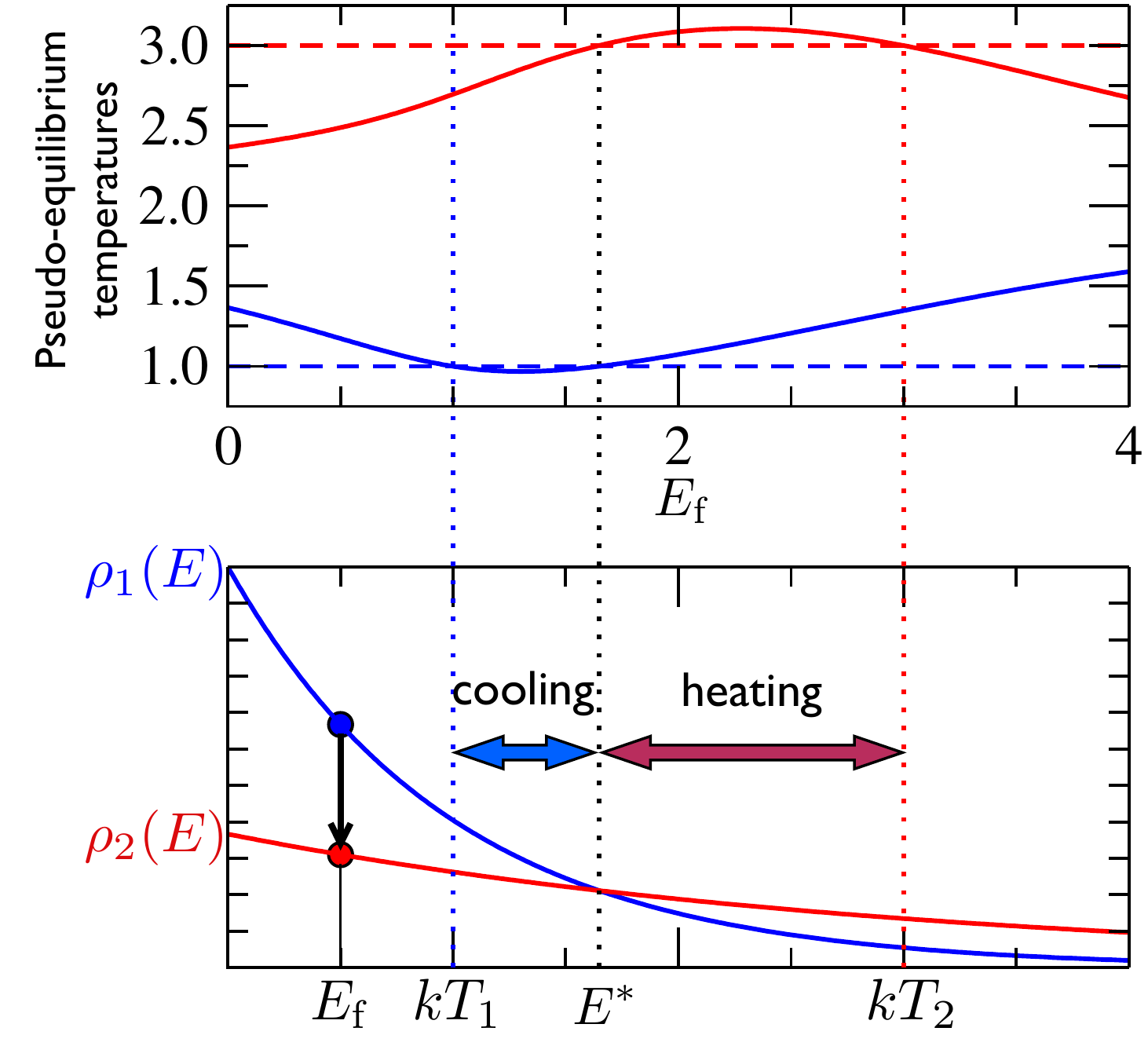}
\caption{ {\em Top:} Pseudo-equilibrium temperatures, $kT_1$ (red) and $kT_2$ (blue),  as a
function of the energy of the filter $E_{\rm f}$, for  initial temperatures and populations $kT_1(0)=1$,
$kT_2(0)=3$, and $N_1(0)=N_2(0)=1000$. The horizontal dashed lines represent
the initial temperatures. The vertical lines depict the  initial average
energies $kT_2(0)=1$ (red), $kT_1(0)=3$ (blue) and the crossing point of
the two Maxwellian distributions $E^*=1.8484\dots$
(black). {\em Bottom:}  Initial Maxwellian distributions $\rho_i(E)$ of 2-d gases with $kT_2(0)=3$ and $kT_1(0)=1$. Particles move from the gas with the higher value of the  Maxwellian at the energy filter as indicated by the vertical arrow for a specific case with $E_{\rm f}<kT_1$. Depending on the location of the filter $E_{\rm f}$ both systems simultanously cool down (blue horizontal arrow) or heat up (red horizontal arrow).}
 \label{fig1}
\end{center}
\end{figure}

We now proceed to characterize the pseudo-equilibrium state given by the condition $\Delta \alpha=0$. As already mentioned, this state exhibits some amusing properties which can be
illustrated in the simple case of classical ideal gases. From here on, we 
consider ideal gases confined in a two-dimensional volume $V$ in order to compare with numerical
simulations. In that case, the equilibrium condition $\Delta \alpha=0$ reduces to Eq.~\eqref{eqgi},
  which has a simple physical interpretation.
Each side of Eq.~\eqref{eqgi},  $\rho_i(E_{\rm f})\equiv\frac{n_i}{kT_i}e^{-E_{\rm f}/kT_i}$, is  the Maxwellian distribution times the density of particles, i.e., is the number of particles with energy $E_{\rm f}$ per unit volume in gas $i$ (for simplicity we will refer to the distribution $\rho_i(E_{\rm f})$ as the Maxwellian of gas $i$). The equilibrium condition then requires that the two distributions intersect at the filter energy $E_{\rm f}$ \cite{linke,linketherm}.

Equation \eqref{eqgi}, together with the state equation $kT_i=E_i/N_i$ and  the constraints $N_1+N_2=N_1(0)+N_2(0)$, $E_1+E_2=E_1(0)+E_2(0)$, $E_1-E_1(0)=E_{\rm f}[N_1-N_1(0)]$, can be solved in terms of the initial number of particles 
and initial energies. The solution univocally determines the pseudo-equilibrium state, i.e., the number of particles and the energy in each gas. Here we discuss instead the temperatures $T_i=E_i/(kN_i)$, which exhibit the most striking features.

Figure \ref{fig1} (top) shows the pseudo-equilibrium temperatures as a function of the energy of the filter $E_{\rm f}$ for the hot (red solid line) and cold (blue solid line) gas with equal initial densities $N_1(0)=N_2(0)$ and initial temperatures $kT_1(0)=1$ (blue dashed line)  and $kT_2(0)=3$ (red dashed line). 
 The figure reveals a rich and counterintuitive behaviour. For either $E_{\rm f}\leq kT_1(0)=1$ or $E_{\rm f}\geq kT_2(0)=3$,  the temperature of the cold gas increases and the temperature of the hot gas decreases, as expected. On the other hand, for $E_{\rm f}$ between the average energy of the cold gas, $kT_1(0)=1$, and $E^*=1.8484...$ 
the temperature of the two gases decreases simultaneously. Finally, for $E_{\rm f}$ between $E^*$ and $kT_2(0)=3$ the two gases heat up. 
These two regimes, simultaneous cooling and simultaneous heating, are in apparent contradiction with our most basic thermodynamic intuition.
Despite their oddness, the spontaneous relaxation to all those pseudo-equilibrium states does not violate the second law of thermodynamics, since the evolution, governed 
by Eq.~\eqref{pheno}, yields $\dot S=\kappa \Delta\alpha^2>0$.

One can get an intuitive picture of the behaviour of the two gases by considering their initial
Maxwellian distributions, $\rho_1(E)$ and $\rho_2(E)$ for $kT_2=3$ and $kT_1=1$, see Fig.~\ref{fig1} (bottom).
The plot shows that the point $E^*$ 
separating the region of simultaneous cooling from
the region of simultaneous heating is precisely the energy where the two Maxwellian distributions 
intersect. The location of the filter with respect to this point $E^*$ determines the direction 
of the net flow of particles in the system. According to Eqs.~\eqref{linearfinaln} and \eqref{f2d}, the flow always goes from the gas with the higher Maxwellian 
distribution at $E_{\rm f}$, as sketched by the two circles in the figure. Then, if $E_{\rm f}< E^*$ 
the net flow of particles goes from gas 1 to gas 2, whereas if the filter energy is above $E^*$ 
particles will be transferred from gas 2 to gas 1.

Now it is clear why both gases increase their temperatures when  $kT_1(0)<E^*<E_{\rm f}<kT_2(0)$: particles with an energy below the average energy of gas 2 and above the average of gas 1 are transferred from gas 2 to 1, hence the 
average of {\em both} gases increase.

At first sight, one could suspect that  a Maxwell demon is needed to carry out this selective transfer of particles. However, this is not the case: passive energy filters are perfectly compatible with thermodynamics since they do not break detailed balance and, as we have already shown, ``heating without heat"  is accompanied by an increase of entropy. The simultaneous cooling can be explained in similar terms.

We can have even richer scenarios if the crossing energy $E^*$ is not located in between $kT_1(0)$ and $kT_2(0)$ or if 
the two initial Maxwellians do not intersect (these situations can occur  if the initial densities are different).  One can e.g 
decrease the temperature of the cold gas and increase that of the hot gas if $E^*< E_{\rm f}< kT_1(0)<kT_2(0)$.
Using the appropriate state equation for the chemical potential, one can prove that the above results are valid for hard disks at high density. Recall that, in this case, the temperature is still given by $T_i= E_i/(kN_i)$.

One of our basic assumptions is to consider the two gases in thermal equilibrium along the whole process. In real situations, however, the exchange of particles occurs in certain region and could induce inhomogeneities in the two systems. 
To study whether these inhomogeneities could affect our results and analyze the aforementioned separation of time scales, we have carried out molecular dynamic simulations of two gases composed by hard disks, placed in two 
square compartments of size $L\times L$. The whole wall separating the two compartments acts as a effusion filter of energy $E_{\rm f}$ and width $\Delta E$.

We place $N_i(0)$ particles 
with a Maxwellian  velocity distribution 
at temperature $T_i(0)$ and implement an
event driven dynamics: 
free motion and collisions conserving energy and momentum. 
When a particle reaches the filter, it can cross it if its energy is between $[E_{\rm f}-\Delta E/2,E_{\rm f}+\Delta E/2]$. While a particle is crossing the filter, 
no collisions are allowed.
If the energy is outside  the interval, particles undergo elastic collisions with the  wall
(adiabatic wall).

\begin{figure}
\begin{center}
\includegraphics[width=0.48\textwidth]{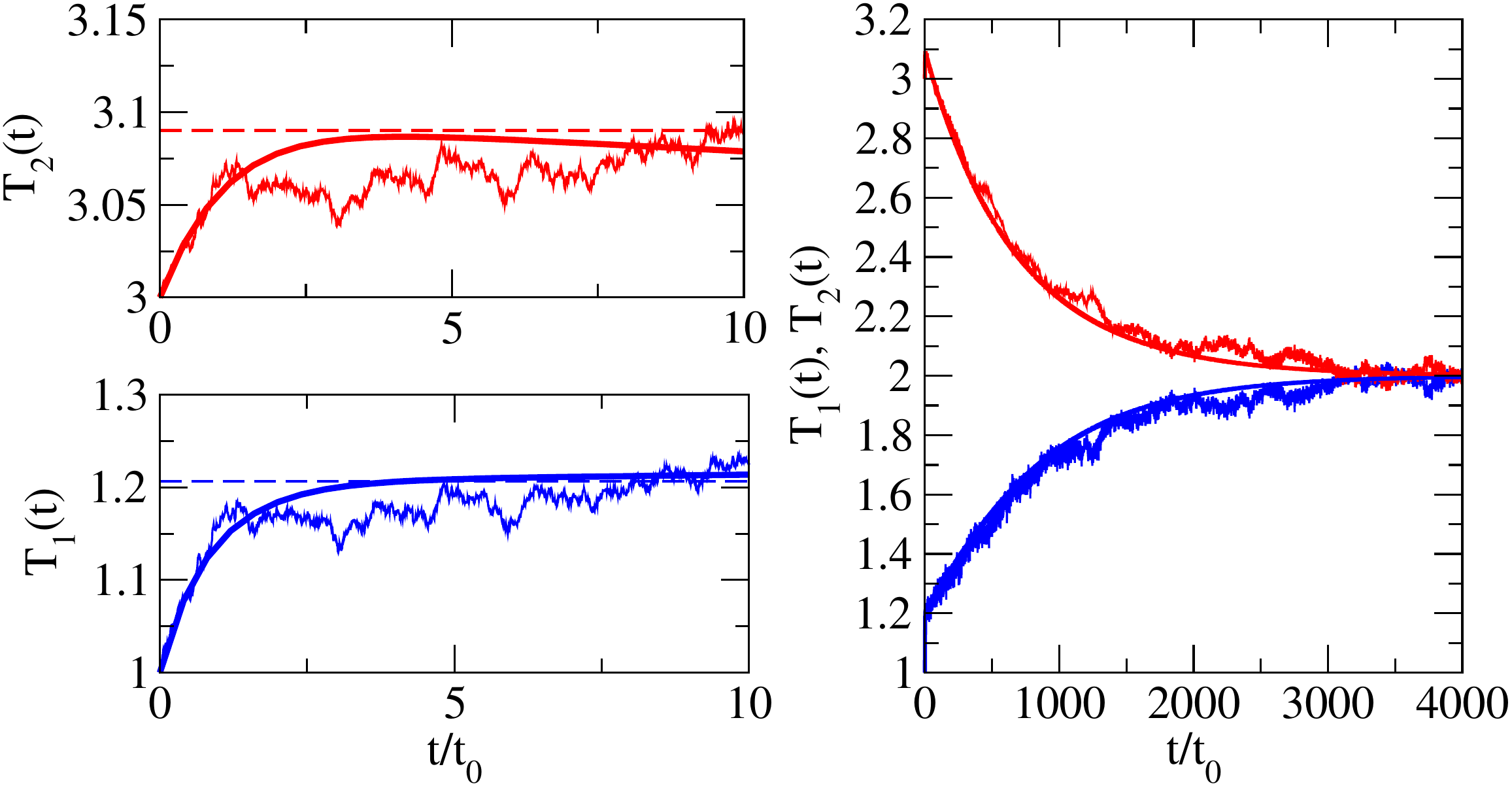}
\caption{Numerical evolution of the temperatures $T_1(t)$ (blue) and $T_2(t)$ (red) as a
function of reduced time $t/t_0$ for a single realization of two-dimensional gases of hard disks with mass $m=1$ and diameter $\sigma=1$. The relaxation time $t_0$ is obtained analytically by expanding Eq.~\eqref{eq1} as $\dot N_1(t)=(N_1(t)-N'_1)/t_0+\dots$ where $N'_1$ is the number of particles in gas 1 in the pseudo-equilibrium state. The rest of parameters are: $E_{\rm f}=2.5$, $\Delta E=0.3$, $k=1$, $T_1(0)=1$,
$T_2(0)=3$, $N_1(0)=N_2(0)=1000$ and $L=200$. 
Solid lines are the solution of Eqs.~\protect{(\ref{eq1})} and \protect{(\ref{eq2})}. {\em Left:} Short time evolution, where the system relaxes to the pseudo-equilibrium state (dashed lines indicate the pseudo-equilibrium temperatures). {\em Right:} Long time evolution, where the system relaxes to full equilibrium. Note the difference in time scales.}
 \label{figsuben}
\end{center}
\end{figure}

Figure \ref{figsuben} (left) shows a case where 
both gases heat up without any energy supply. 
This happens in the short time range, where the filter behaves as a perfect one and only the first term in (\ref{linearfinaln}-\ref{linearfinale}) describes the dynamics of the system. 
The long time behaviour
is presented in  
Fig.~\ref{figsuben} (right). 
The system decays to a truly thermodynamic equilibrium in a time about three orders of magnitude longer, in agreement with our previous analysis. The smooth solid lines depict the exact solutions of Eqs.~(\ref{eq1}-\ref{eq2}). The good agreement between the simulations and the evolution predicted by kinetic theory indicates that the inhomogeneities do not play a relevant role.

\begin{figure}[h]
\begin{center}
 \includegraphics[width=0.45\textwidth]{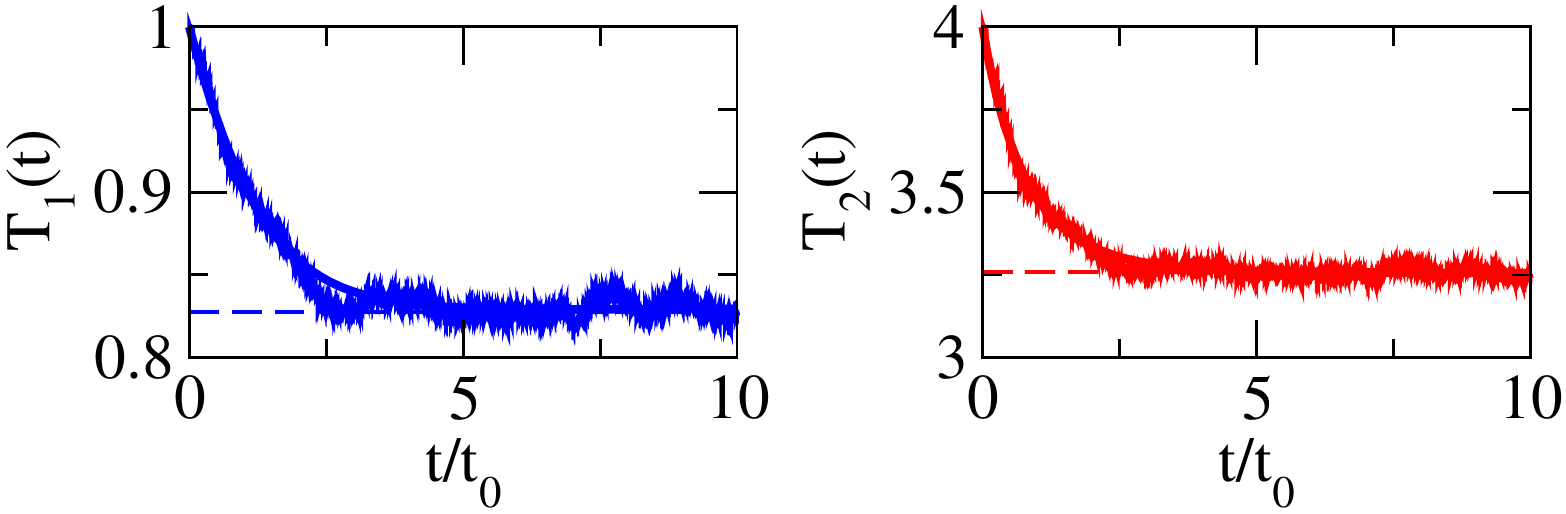}\\[0.4cm]
\includegraphics[width=0.45\textwidth]{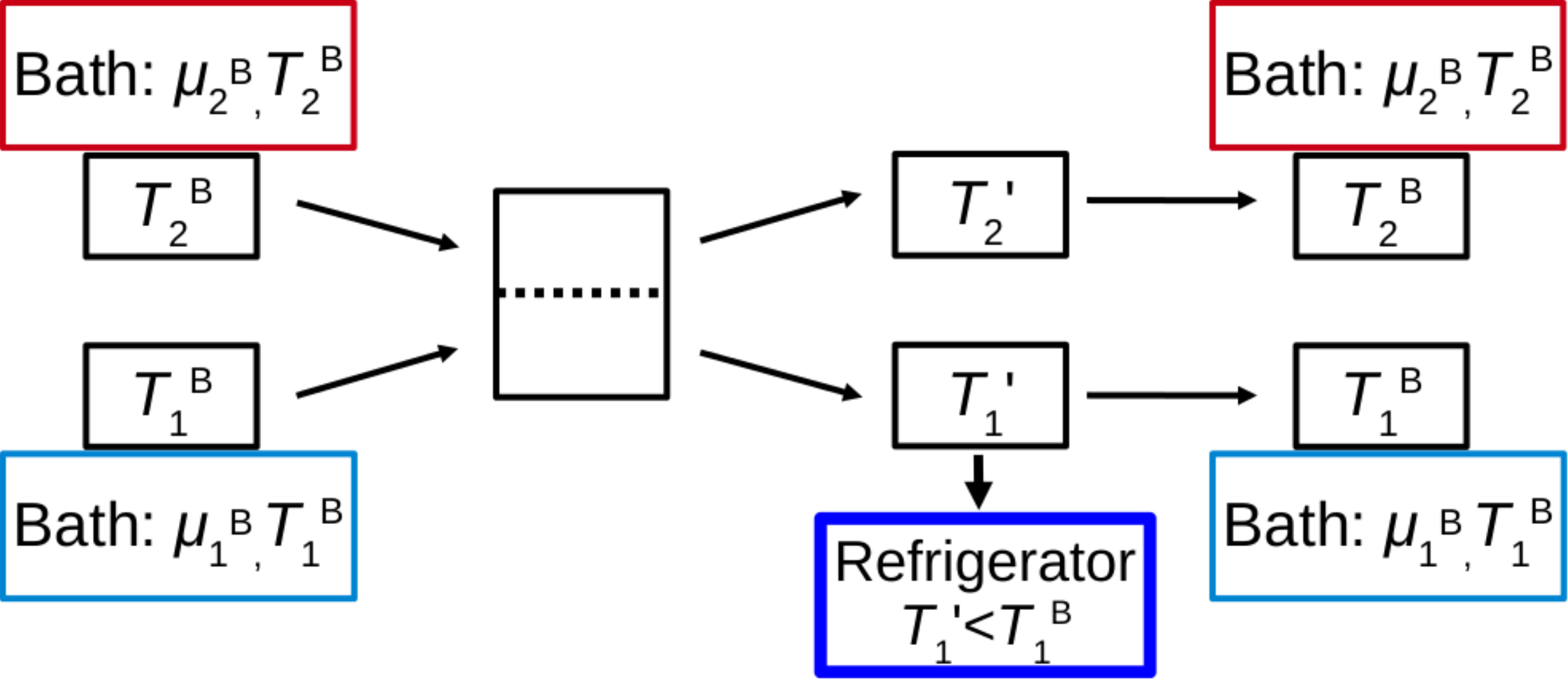}
\caption{
{\em Top:} Evolution of the temperatures $T_1$ (blue) and $T_2$ (red) as a
function of reduced time $t/t_0$ for a case where both gases cool down. Parameters are: 
$E_{\rm f}=1.6$, $\Delta E=0.1$, $k=1$, $T_1(0)=1$,
$T_2(0)=4$, and $N_1(0)=4000, N_2(0)=2000$ and $L=400$. {\em Bottom:} Implementation of a cycle where, by using an energy filter, we can  lower the temperature of a system (and transfer it to a refrigerator) below the temperature of both thermal reservoirs without energy consumption or mechanical device.}
 \label{figbajan}
\end{center}
\end{figure}

The remarkable phenomena discussed along this Letter could be exploited to design interesting setups with potential applications. For instance, one can use thermal reservoirs to reset the systems to their initial temperatures and densities. 
We show in Fig.~\ref{figbajan} a cycle built upon this idea. The two systems are represented by black small boxes and the cycle starts with each of them exchanging particles and energy with a suitable reservoir. In this step, the exchange is not restricted by any filter, so each pair, system and reservoir, reaches full equilibrium with equal temperatures,  $T^{\rm B}_1$ and  $T^{\rm B}_2$, and chemical potentials, $\mu^{\rm B}_1$ and  $\mu^{\rm B}_2$.  Suppose that the temperature and chemical potential of the reservoirs are such that the two systems equilibrate at the initial state of Fig.~\ref{figbajan} (top), given by $kT_1(0)=kT_1^{\rm B}=1$, $kT_2(0)=kT_2^{\rm B}=4$, $N_1(0)=4000$ and $N_2(0)=2000$. In the next step,  the two systems are connected by a filter centered at $E_{\rm f}=1.6$, exactly as in Fig.~\ref{figbajan} (top), and therefore, reach temperatures $T'_2<T_2^{\rm B}$ and $T_1'<T_1^{\rm B}$. The cold system 1 can now be used to refrigerate a third system down to $T'_1\simeq 0.83$. Finally, the cycle is closed by connecting again the two systems with their respective reservoirs. The cycle is of course irreversible. Nevertheless, it is remarkable that we obtain an effective reservoir of temperature $T'_1$, lower than the temperature of the two reservoirs 1 and 2. The cold system at temperature $T_1'$ can be considered  an effective reservoir since the cycle can be repeated as often as desired, extracting or releasing heat but keeping its temperature equal to $T'_1$ in the refrigeration step. Notice that this effective reservoir at temperature $T_1'<T_1^{\rm B}<T_2^{\rm B}$ is obtained without the need of any mechanical work.

To conclude, we have shown that the selective exchange of particles between finite systems induces a rich and counterintuitive thermodynamic behaviour. Some of the induced phenomena seem to defeat our most basic intuitions on how temperature changes and heat flows in isolated  systems. Yet, they can be understood and analyzed using elementary kinetic and thermodynamic arguments. 
Our results are of theoretical relevance, since they illustrate that
equilibrium is not restricted to the standard repertoire of thermal (equal temperatures), chemical (equal chemical potentials) and mechanical (equal pressures) equilibrium. Moreover, the ideas in this Letter  could be useful to design thermostats and calorimetric devices, as the one sketched in Fig.~\ref{figbajan}.
However, in this respect it is worth mentioning two difficulties. First,   simultaneous cooling or heating is much harder to be found in degenerate Fermi gases ($n\Lambda^d \gtrsim 1$,  with $d$ the dimension). 
The main reason is that energy is not as sensitive to temperature as it is in Bose systems or in non-degenerate gases ($n\Lambda^d \ll 1$), like the one analyzed here. Second, up to our knowledge, energy filters for classical particles have not been developed so far, although  there is no any fundamental reason to prevent their existence. Therefore, the implementation of the scenarios discussed above in real settings, with either electrons or
classical particles, is a particularly challenging endeavour.

\begin{acknowledgments}
 R.M.T acknowledges 
 Spanish MINECO  grants FIS2013-40627-P and FIS2016-80681-P (AEI/FEDER, UE), and  Generalitat de Catalunya CIRIT  2014-SGR-966.
R.B. and J.M.R.P. acknowledge financial support from MINECO grant FIS2014-52486-R.
\end{acknowledgments}


\begin{thebibliography}{0}
\bibitem{finiteness} Our results would also apply to infinite systems ($N\to\infty$) if the flow of particles and energy between the two systems were proportional to $N$. This situation is rather unlikely. For three (two) dimensional gases
 the flow is, in most cases, proportional to the contact area (length) between the two gases, i.e., the flow is proportional to $N^{2/3}$($N^{1/2}$).  A counterexample, however, is given by two dimensional gases exchanging particles through a two dimensional film. 
We prefer to keep the term ``finite'' to stress that our results do not apply to reservoirs.
\end{thebibliography}
\end{document}